\documentclass[a4paper,twoside]{article}

\usepackage{epsfig}
\usepackage{subcaption}
\usepackage{calc}
\usepackage{amssymb}
\usepackage{amstext}
\usepackage{amsmath}
\usepackage{amsthm}
\usepackage{multicol}
\usepackage{pslatex}
\usepackage{apalike}
\usepackage{algorithm2e}
\usepackage[bottom]{footmisc}
\usepackage{SCITEPRESS}     
\usepackage{tabularx}
\usepackage{cite}
\usepackage{comment}
\usepackage{orcidlink}
\usepackage{verbatim}
\usepackage{textcomp}
\usepackage{blindtext}
\usepackage{nicefrac}
\usepackage{stfloats}

\usepackage{cite}
\usepackage{amsmath,amssymb,amsfonts}
\usepackage{algorithmic}
\usepackage{graphicx}
\usepackage{textcomp}
\usepackage{xcolor}
\usepackage{tikz}
\usetikzlibrary{quantikz2}
\usepackage{bm}  
\usepackage{caption}
\usepackage{comment}
\usepackage{appendix}

\def\BibTeX{{\rm B\kern-.05em{\sc i\kern-.025em b}\kern-.08em
T\kern-.1667em\lower.7ex\hbox{E}\kern-.125emX}}

\begin{document}

\title{Accelerated VQE: Parameter Recycling for Similar Recurring Problem Instances}

\author{\authorname{Tobias Rohe\sup{1}\orcidlink{0009-0003-3283-0586}, Maximilian Balthasar Mansky\sup{1}, Michael Kölle\sup{1}\orcidlink{0000-0002-8472-9944}, Jonas Stein\sup{1,2}\orcidlink{0000-0001-5727-9151}, Leo Sünkel\sup{1}\orcidlink{0009-0001-3338-7681}, and Claudia Linnhoff-Popien\sup{1}\orcidlink{0000-0001-6284-9286}}
\affiliation{\sup{1}Mobile and Distributed Systems Group, LMU Munich, Germany}
\affiliation{\sup{2}Aqarios GmbH, Germany}
\email{tobias.rohe@ifi.lmu.de}
}


\abstract{Training the Variational Quantum Eigensolver (VQE) is a task that requires substantial compute. We propose the use of concepts from transfer learning to considerably reduce the training time when solving similar problem instances. We demonstrate that its utilisation leads to accelerated convergence and provides a similar quality of results compared to circuits with parameters initialised around zero. Further, transfer learning works better when the distance between the source-solution is close to that of the target-solution. Based on these findings, we present an accelerated VQE approach tested on the MaxCut problem with a problem size of 12 nodes solved with two different circuits utilised. We compare our results against a random baseline and non transfer learning trained circuits. Our experiments demonstrate that transfer learning can reduce training time by around 93\% in post-training, relative to identical circuits without the use of transfer learning. The accelerated VQE approach beats the standard approach by seven, respectively nine percentage points in terms of solution quality, if the early-stopping is considered. In settings where time-to-solution or computational costs are critical, this approach provides a significant advantage, having an improved trade-off between training effort and solution quality. }

\onecolumn \maketitle \normalsize \setcounter{footnote}{0} \vfill

\section{Introduction}\label{sec:introduction}

Quantum computing (QC) has applications in chemical simulations \cite{kassal2011simulating}, machine learning \cite{schuld2015introduction}, and optimization \cite{moll2018quantum}. Despite hardware limitations \cite{lubinski2023application}, the quantum approximate optimization algorithm (QAOA) \cite{farhi2014quantum} and variational quantum eigensolver (VQE) \cite{peruzzo2014variational} are promising for noise resilience. While the QAOA is often used for combinatorial optimisation~\cite{khairy2020learning, blekos2024review, cheng2024quantum} and the VQE often for chemical problems~\cite{lee2018generalized, grimsley2019adaptive, blunt2022perspective, fedorov2022vqe}, also the VQE shows potential for combinatorial optimization (CO) problems \cite{amaro2022filtering, kolotouros2022evolving, niroula2022constrained, liu2022layer}, offering lower circuit depth beneficial in the NISQ era \cite{preskill2018quantum}. Nevertheless, training VQE can be challenging due to issues like barren plateaus \cite{liu2024mitigating}.

This paper explores using transfer learning (TL) with VQE for CO problems, a method yet to be investigated \cite{truger2024warm}. We focus on problems like the capacitated vehicle routing problem (CVRP)~\cite{ralphs2003capacitated}, which often demands for quick, efficient solutions for similar problem instances in logistics \cite{fukasawa2006robust, gonzalez2008models}. Our study assesses TL's effectiveness and the application of an accelerated VQE process, which reduces training time by 93\% while maintaining high solution quality. This approach is advantageous where solutions need rapid, cost-effective resolution and problem instances are similar, whereby we adhere to the general quantum development pipeline~\cite{rohe2025problem}.

The paper is structured as follows: Sec. \ref{sec:background} \textit{Background} outlines foundational concepts, Sec. \ref{sec:method} \textit{Methodology} details our algorithms, Sec. \ref{sec:results} \textit{Results} presents our findings, Sec. \ref{sec:accelerated-vqe} \textit{Accelerated VQE} explores the TL benefits for quick parameter training, and Sec. \ref{sec:discussionlimitations} \textit{Discussion \& Limitations} examines the study's implications and limits. Sec. \ref{sec:conclusion} \textit{Conclusion} encapsulates the research's main points.

\section{Background}\label{sec:background}

\subsection{MaxCut Problem}
The maximum cut (MaxCut) problem asks for a binary partition of a graphs $n$ vertices, such that the number of edges between the vertices of both partitions is maximised. Identifying the optimal partition for a general graph is NP-hard. We formally write the graph $G = (V, E)$ with $n$ nodes $V = \{0,...,n-1\}$ and undirected edges $\{i, j\} \in E$. In our case, the graph is unweighted, each edge has the same weight of $w_{i,j} = 1$. The partitioning of the nodes into the two sets is encoded in binary as $z_i= 1$ for a node $i$ being contained in the first set, respectively $z_i= -1$ for the second set. The corresponding cost function, representing the number of edges that are cut by the partition, is given by:

\begin{equation}
C(z) = \sum_{(i, j) \in E} \frac{1 - z_i z_j}{2}
\end{equation}

This cost function evaluates the total number of edges that span between the two partitions, thus representing the MaxCut problem. In the following sections, we reformulate this maximization problem as an equivalent minimization problem to leverage specific optimization techniques.

\subsection{Variational Quantum Eigensolver}
The VQE is a hybrid quantum-classical algorithm, composed of both a classical and a quantum part, applied in a loop. It was first proposed by Peruzzo et al. \cite{peruzzo2014variational} and is in broad usage since then \cite{cerezo2022variational, tilly2022variational, kandala2017hardware, wang2019accelerated}. The quantum part of the algorithm consists of a parameterised quantum circuit (PQC), also known as Ansatz, with parameters $\theta\in \mathbb{R}^m$, described by $f:\theta \mapsto U\left(\theta\right)\ket{0}^{\otimes n}$, preparing the quantum state $\ket{\psi\left(\theta\right)}\coloneqq U\left(\theta\right)\ket{0}^{\otimes n}$. The prepared quantum state represents the expectation value of the problem Hamiltonian $\hat{H}$, where the problem Hamiltonian can be  derived from the underlying problem's cost function. The ground state of the Hamiltonian then represents the optimal solution of the corresponding cost function. This is done by iteratively optimising the gate parameters $\theta$ of the PQC with a classical optimiser in the classical part of the algorithm. The quantum-classical loop is closed after the optimised parameters are returned to the quantum part and the PQC is executed again. The algorithm's execution ends as soon as a predefined convergence criterion is reached, e.g. the expectation value changes are minimal, or the maximum number of loop iterations (\texttt{maxiter}) has been reached.

\subsection{Transfer Learning in Quantum Computing}
While TL is a well-established technique in classical machine learning~\cite{zhuang2020comprehensive}, in quantum machine learning it is just emerging. TL involves the practice of applying pre-trained knowledge, corresponding parameters and weights, to a new, potentially related, task and/or dataset~\cite{mari2020transfer}. This approach allows training to commence at a higher accuracy level, leveraging the relevance of the transferred parameters to the problem class. Consequently, for an equivalent number of training steps, a higher accuracy is generally achievable~\cite{torrey2010transfer}.

In the context of QC and TL, an early contribution was made by Brandao et al. in 2018~\cite{brandao2018fixed}, who explored the QAOA for the MaxCut problem on 3-regular graphs. Their research unveiled that QAOA parameters trained for a random instance also work well for related instances, sampled from the same underlying distribution. In practical terms, once optimal parameters are determined for one instance, these parameters also yield effective results for other similar instances, leading to a significant reduction in the computational overhead for subsequent instances as these pre-trained parameters can be re-used for another QAOA instance~\cite{brandao2018fixed}. Furthermore, their work posits the feasibility of transferring parameters across problems of varying sizes, especially those requiring more extensive circuits. They call it ``leapfrogging'' method~\cite{brandao2018fixed}, where parameters optimised for smaller instances can be used as starting points for larger instances. This approach effectively sidesteps the exhaustive search for optimal parameters for every new, but similar problem instance. 

Further researchers~\cite{galda2021transferability, galda2023similarity} have investigated the transferability of QAOA parameters for 3-regular graphs and broadly confirm the results of Brandao et al.~\cite{brandao2018fixed}. They highlight the importance of similarity in the instances subgraph decomposition, therefore their local properties, as major factor for the successful application of parameter transfer. These findings are complemented by the work of Wurtz and Love~\cite{wurtz2021maxcut}, who provide worst case lower bound performance guarantees for uniform 3-regular graphs at particular fixed parameters for $p = 2$ and $p = 3$, as well as upper bounds for all $p$`s. For this so-called fixed angle conjecture numerical evidence for $p < 12$ is provided here~\cite{wurtz2021fixed}.    

Beside these major contributions, other scientific papers have researched parameter transfer for QAOA, highlighting stable or even improved objective values while drastically reducing training times~\cite{shaydulin2019multistart, shaydulin2023parameter,liu2023mitigating, montanez2024transfer}. A dedicated framework for evaluation is available, containing pre-optimised QAOA parameters and circuits for different instances~\cite{shaydulin2021qaoakit}. Furthermore, pre-optimised QAOA parameters, initially trained on unweighted graphs, can be efficiently applied to weighted graphs upon appropriate rescaling~\cite{sureshbabu2024parameter}. 

TL has not been widely applied to VQE algorithms~\cite{skogh2023accelerating}, however, the idea has been proposed before ~\cite{shaydulin2019multistart}. To the best of our knowledge only one paper specifically covers this area of research, while focusing on potential energy surface calculations for molecules~\cite{skogh2023accelerating}. The researchers aim at a faster convergence towards the ground state of the target molecule through initialising the VQE algorithm with the pre-calculated ground state of molecules with similar geometry. The results indicate speedup potential with regards to standard initialisation, on condition that a suitable optimisation algorithm was used. 

A related concept is the so-called warm-starting technique, which also comes from the field of classical machine learning and optimisation. Warm-starting uses approximated solutions for initialisation to improve the performance of the algorithm used~\cite{truger2024warm}. This is particularly attractive for problems where a good approximation can be efficiently obtained~\cite{egger2021warm, okada2024systematic, tao2023laws, jain2022graph}.

\section{Methodology}\label{sec:method}
\begin{figure*}[!t]
\centering
\includegraphics[width=\textwidth]{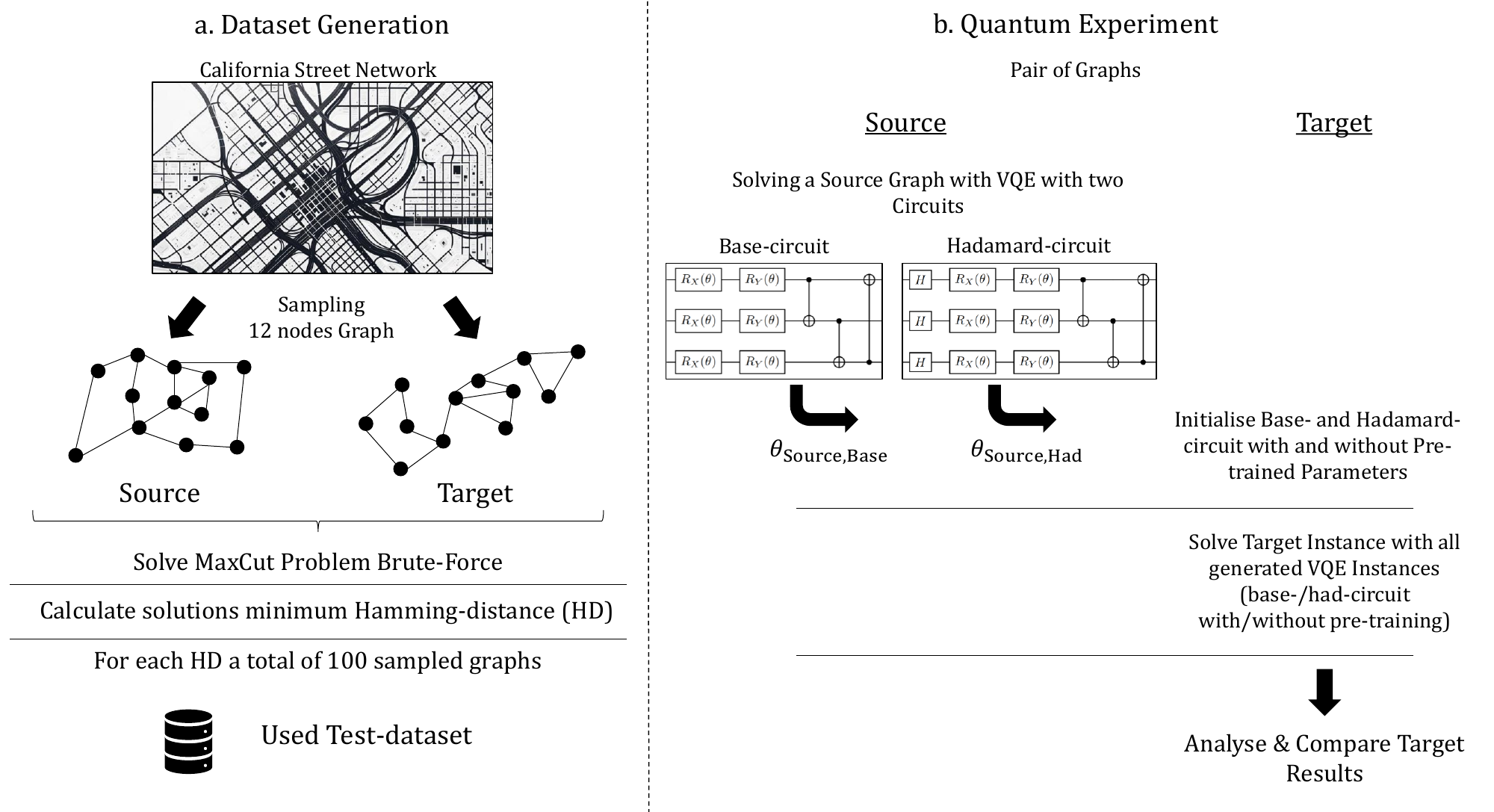}
\caption{Overview of the Research Methodology. Panel a. on the left illustrates the Dataset Generation phase, Panel b) on the right depicts the Algorithmic Setup.
}
\label{fig:method_vis}
\end{figure*}

In the following, we show the methodology, with focus on the dataset creation and the algorithmic setup applied. This is visualised in Figure \ref{fig:method_vis}.

\subsection{Dataset Generation}
We generated a dataset for parameter pre-training and TL evaluation using the MaxCut problem on undirected, unweighted graphs derived from California's road network, which includes about 1.9 million nodes and 2.7 million edges. Our dataset construction involves sampling graph pairs—source and target—with each graph consisting of 12 nodes selected through a neighborhood expansion algorithm ensuring all nodes and edges are unique and connected.

For each graph pair, we identify optimal MaxCut solutions using brute-force, then calculate the smallest Hamming distance (HD) across all optimal pairings, including symmetrical solutions. HD, ranging from 0 to 6, measures the bit flips required to convert one graph's solution to another's, serving as a metric for solution distance. We sample 10 graph pairs for each HD value to ensure a balanced dataset, facilitating the evaluation of source-target graph distance impacts in our TL study.

\subsection{Algorithmic Setup}
\begin{figure}
    \centering
    \resizebox{\columnwidth}{!}{ 
    \begin{quantikz}
        \lstick{$\ket{q_0}$} & \gate{H} \slice{} & \gate{R_X(\bm{\theta_0})} & \gate{R_Y(\bm{\theta_3})} & \ctrl{1} & \qw & \targ{} & \qw \\
        \lstick{$\ket{q_1}$} & \gate{H} & \gate{R_X(\bm{\theta_1})} & \gate{R_Y(\bm{\theta_4})} & \targ{} & \ctrl{1} & \qw & \qw \\
        \lstick{$\ket{q_2}$} & \gate{H} & \gate{R_X(\bm{\theta_2})} & \gate{R_Y(\bm{\theta_5})} & \qw & \targ{} & \ctrl{-2} & \qw
    \end{quantikz}
    }
    \caption{Schematic Illustration of the Base- and Had-Circuits. The Hadamard gates are only present in the had-circuit. For illustrative reasons we only print a three qubit circuit here, the expansion towards a 12 qubit circuit is straightforward. A measurement is carried out at the end of each circuit.}
    \label{fig:circuit}
\end{figure}
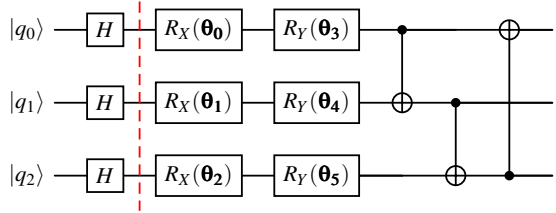

For evaluation, we use two quantum circuits: the base-circuit and the had-circuit, each featuring rotational X-gates and Y-gates followed by a circular CNOT entanglement, differentiated by an initial layer of Hadamard gates in the had-circuit. Both circuits are illustrated in Figure \ref{fig:circuit} and are repeated three times for 12 qubits, serving as a foundation for our proof of concept study, as these architectures can also often be found in literature~\cite{ravi2022vaqem, gocho2023excited, ravi2022cafqa}.

The training consists of two phases. In pre-training, two VQE instances (one with each circuit) are initialised with parameters near zero (range [-0.001, 0.001]~\cite{mcclean2018barren}) and optimised using the COBYLA optimiser~\cite{powell1994direct} with a maximum of 1000 iterations. The optimised parameters, $\theta_{\text{pre}}$, are recorded for the source-graph. Post-training involves using these pre-optimised parameters to initialise the VQE on the target-graph. For comparison, we also train additional VQEs with parameters reinitialised around zero.

We tested our approach with ten different seeds, resulting in a dataset of 700 graph pairs across different HDs, conducted without noise to clearly assess the benefits of our methodology. An additional VQE test with parameters initialised randomly between 0 and 2$\pi$ served as another baseline, with results detailed in Appendix \ref{appendix_a}. This experiment highlights the drawbacks of non-pre-optimised initialisation, as these showed significantly poorer performance, thus omitted from our main findings.

\section{Results}\label{sec:results}

In this section, we first analyse the convergence behaviour, followed by the solution quality obtained through the various training variants. Finally, we will evaluate the performance based on the distance of the solutions from the source- to the target-graphs, measured as the minimum HD of optimal solutions.

\subsection{Convergence Behaviour}

\begin{figure*}[!t]
\centering
\includegraphics[width=\textwidth]{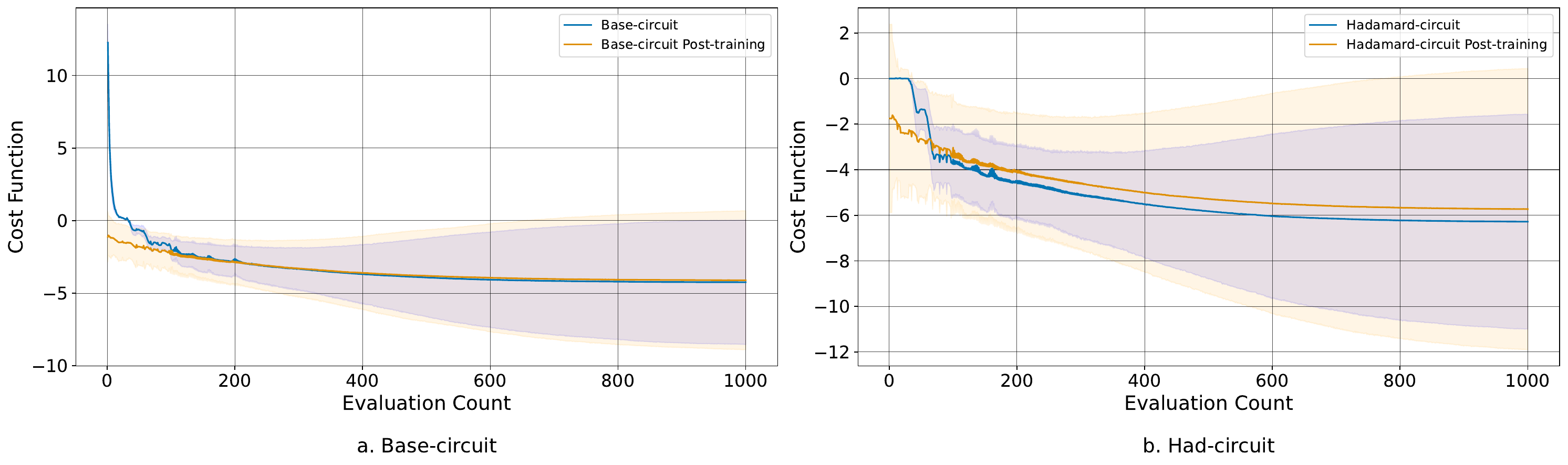}
\caption{Convergence of VQE Implementations (\texttt{maxiter = 1000}). The figure compares the convergence of the a. base-circuit and the b. had-circuit, under two initial conditions: near-zero initialisation (blue) and pre-trained parameters (orange), averaged over all HD. The variance is visualised as a shaded area. Lower values are better.}
\label{fig:combined_convergence}
\end{figure*}

Throughout the execution, we record the convergence behaviour of the different VQE versions, illustrated in Figure \ref{fig:combined_convergence}. The base-circuit with standard initialisation (blue) and the base-circuit with pre-optimised parameter initialisation (orange) is depicted in sub-figure a., while the corresponding data for the had-circuit and its two initialisation variants is presented in sub-figure b.

Observing Figure \ref{fig:combined_convergence} a., we notice that the TL approach curve begins at a significantly lower level compared to the classical initialisation approach. The TL version converges more gradually. Moreover, the convergence threshold is observed to be lower in the classical initialisation version, wherein its variance is marginally larger. 

We interpret this observation as follows: Through the TL, the VQE already represents a state near an optimum of an unknown graph. When this pre-trained instance is applied to a new graph, we can expect that the VQE converges faster since the transferred parameters are closer to a solution of the problem. Rather than starting from scratch, a favourable starting position has already been established, in general, only minor adjustments need to be made to the new solution. Overall, the convergence behaviour is relatively similar.

In Figure \ref{fig:combined_convergence} b., which displays the results for the hadamard-enhanced circuit, we also note a lower starting point for convergence, a more gradual convergence behaviour, and a higher convergence limit for the TL approach. Here too, the variance is found to be lower for the standard training. Interestingly, while the TL approach starts at approximately the same energy level as the base-circuit without a ladder of Hadamard-gates, the convergence curve for the standard learning approach, initialised with gate-parameters around zero, begins significantly lower compared to the base-circuit execution. We explain this by the fact that the Hadamard-gates create a superposition of all nodes, effectively simulating each node's presence in both partitions simultaneously. In contrast, the basic circuit's initialisation around zero begins as if all nodes were in a single group, yielding the worst possible outcome, zero cuts in the graph. Similar to the base circuit described above, the other observed characteristics $-$ a more gradual convergence, increased variance, and higher convergence level for the TL approach $-$ are also present here.

\begin{figure}[!t]
    \centering
    \includegraphics[width=\columnwidth]{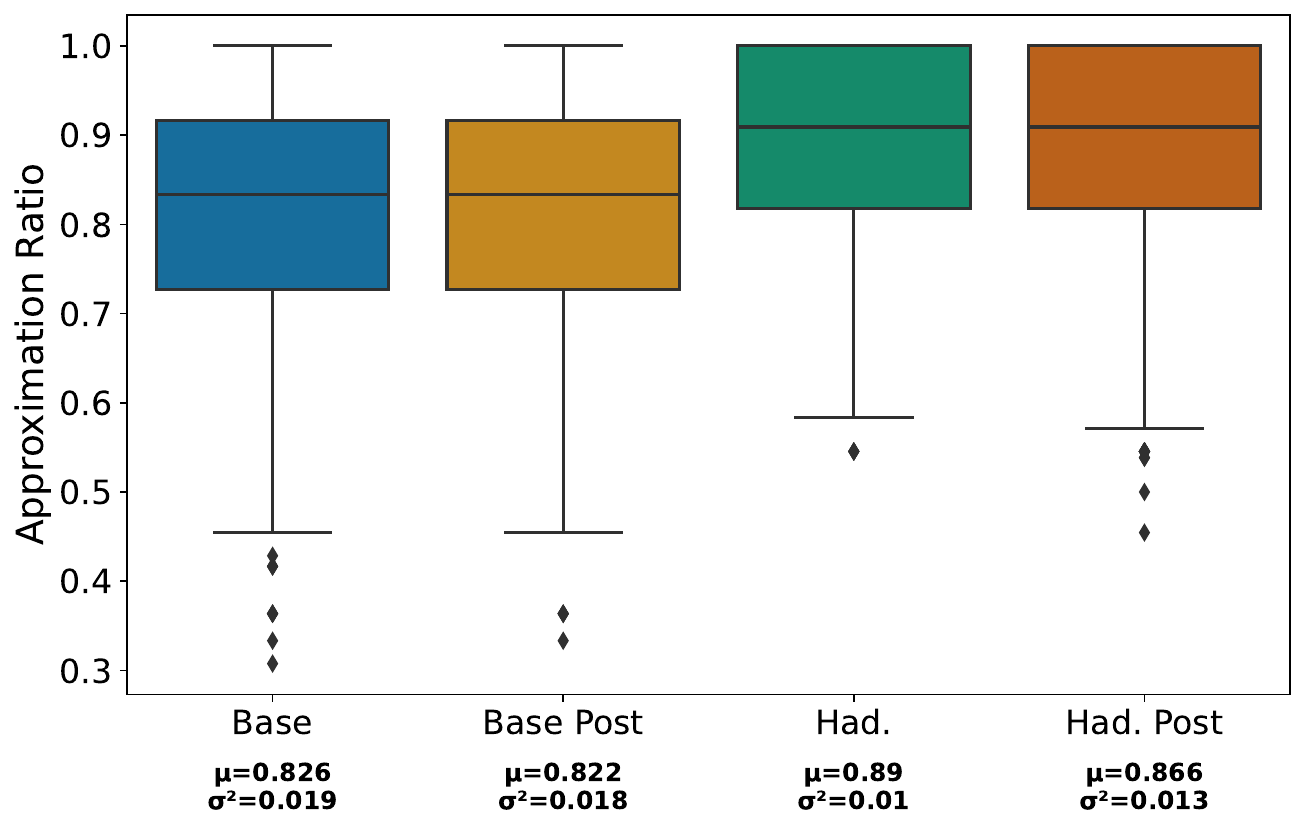}
    \caption{Approximation Ratios of VQE Implementations (\texttt{maxiter = 1000}). This figure displays the approximation ratios for all four versions of the VQE tested. High values indicate better solution quality, with the best achievable ratio being one.}
    \label{fig:boxplot_approxRatios}
\end{figure}

\subsection{Solution Quality}
We display the solution quality of the two different circuit-types with the two distinct training approaches in Figure \ref{fig:boxplot_approxRatios}. The boxplot diagram illustrates the average approximation ratios of the various executed VQE instances. The mean values and variances are shown beneath the corresponding boxplots. Two observations can be made. First, the had-circuit outperforms the base-circuit, achieving a higher approximation ratio as well as exhibiting a lower variance in the observations. This becomes particularly clear when the performance of the two circuit types is compared on the graph-level individually. Here, the had-circuit performs better than the base-circuit in $52.3\%$ and $47.7\%$ of all graphs in classical training and post-training respectively. While the had-circuit only provides worse approximation ratios than the base-circuit in $24.9\%$ and $28.7\%$ of all solved graphs. 
Second, the performance of the two training variants, when compared to each other within the same circuit, is relatively comparable. Although the classical training method slightly performs better then the TL approach, this advantage is marginal and therefore insignificant. It cannot be concluded that there is a significant difference in performance between the two circuit initialisation variants, leading us to consider the quality of the results as comparable.

\subsection{Solution Distance Dependent Analysis}

\begin{figure*}[!t]
\centering
\includegraphics[width=\textwidth]{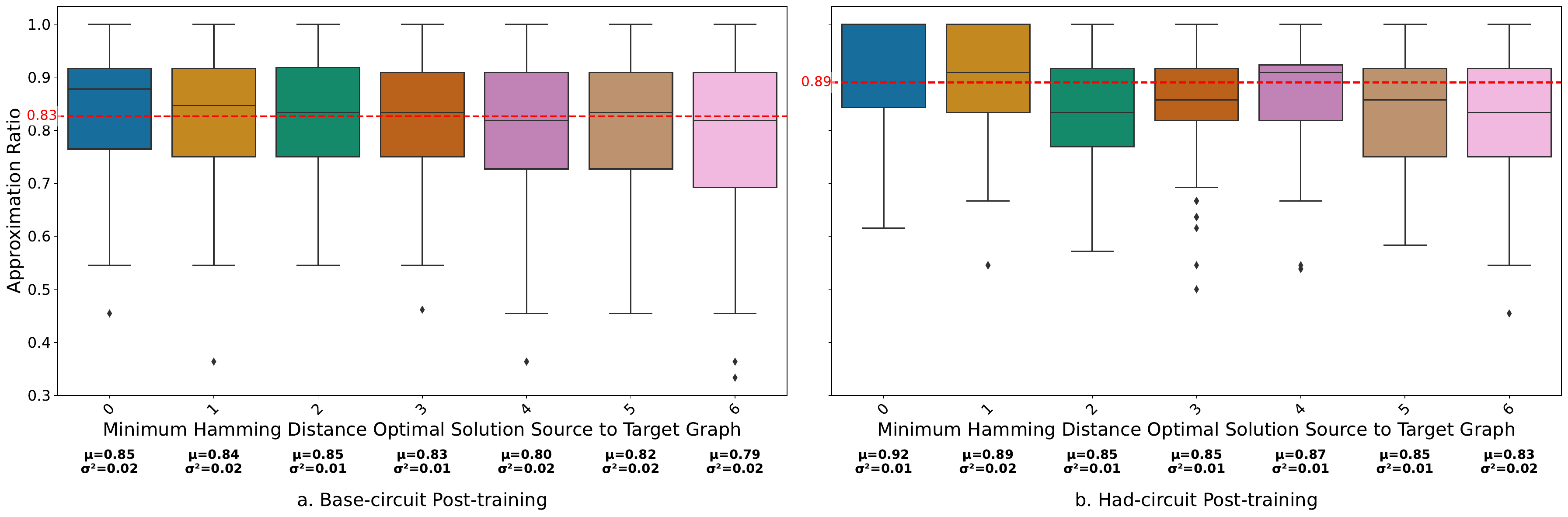}
\caption{Approximation Ratios of the Post-trained VQE instances with regards to the Solutions HD (\texttt{maxiter = 1000}). The figure illustrates the performance of the post-trained VQE instances (a. base-circuit, b. had-circuit) with regards to the minimum HD of the optimal solutions of the underlying source-target graph pairs. High values indicate a better solution quality, with the best achievable ratio being one.}
\label{fig:boxplot_hd}
\end{figure*}

In our study, we assess the impact of the HD between the optimal solutions of source and target graphs on the effectiveness of the TL approach in VQE settings. Our analysis, depicted in Figure \ref{fig:boxplot_hd}, illustrates the approximation ratios for circuits trained with pre-optimised parameters based on the minimum HD between optimal solutions. The data shows that TL-enhanced circuits generally outperform standard initialisation for smaller HDs (HD zero to three), indicating a clearer trend in improved performance for closely related graph instances.\\
For TL applications, smaller HDs correlate with better performance metrics, such as faster convergence and higher solution quality, suggesting that TL is particularly advantageous for problems with similar or closely related instances. These findings, detailed further in Appendix \ref{appendix_b}, underscore the efficiency of the TL approach in achieving rapid convergence and maintaining high solution quality in instances where source and target graphs are similar.\\
It is critical to note that the HD values were computed based on optimal solutions of the graphs rather than the outcomes of the pre-trained instances, highlighting the strategic application of TL where initial conditions closely mimic optimal solutions, thereby enhancing the overall efficacy of the VQE process. This detailed examination substantiates the TL method's applicability to quantum computing scenarios involving repetitive or similar problem instances, advocating for its broader application in quantum optimisations.
\section{Accelerated VQE}\label{sec:accelerated-vqe}

\begin{figure*}[!t]
\centering
\includegraphics[width=\textwidth]{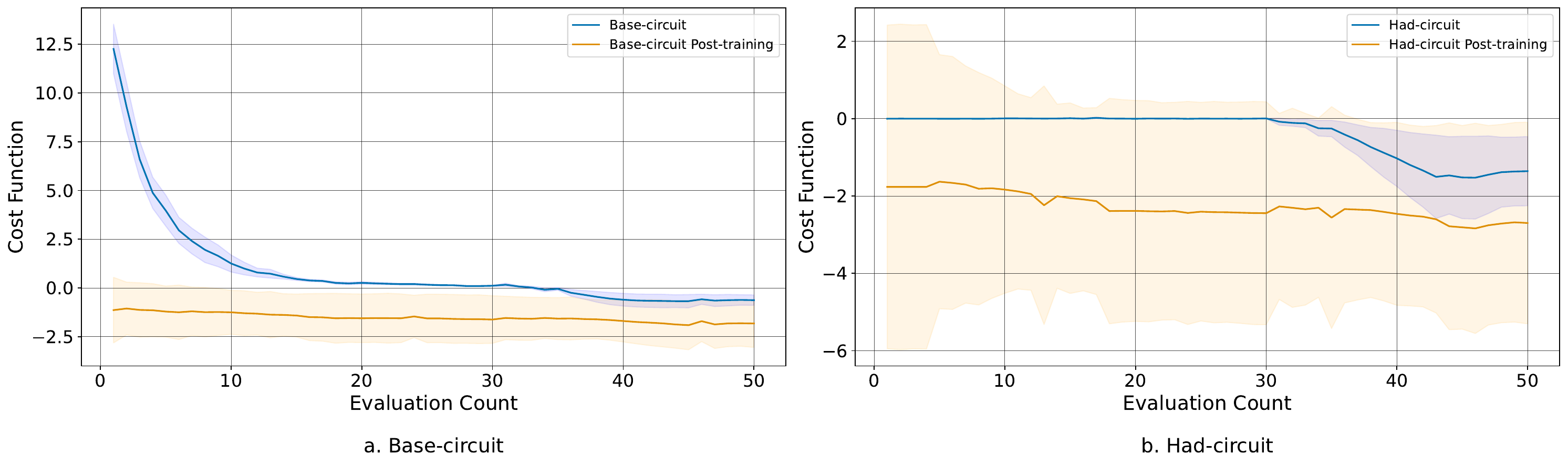}
\caption{Convergence of the accelerated VQE Implementations (\texttt{maxiter = 50}). The figure compares the convergence of the base-circuit (left) and the had-circuit (right), under two initial conditions: near-zero initialisation (blue) and pre-trained parameter initialisation (orange). Here, the \texttt{maxiter} variable was set to 50 iterations. The variance is visualised as a shaded area. Lower values are better.}
\label{fig:combined_convergence_50}
\end{figure*}

\begin{figure}[!t]
    \centering
    \includegraphics[width=\columnwidth]{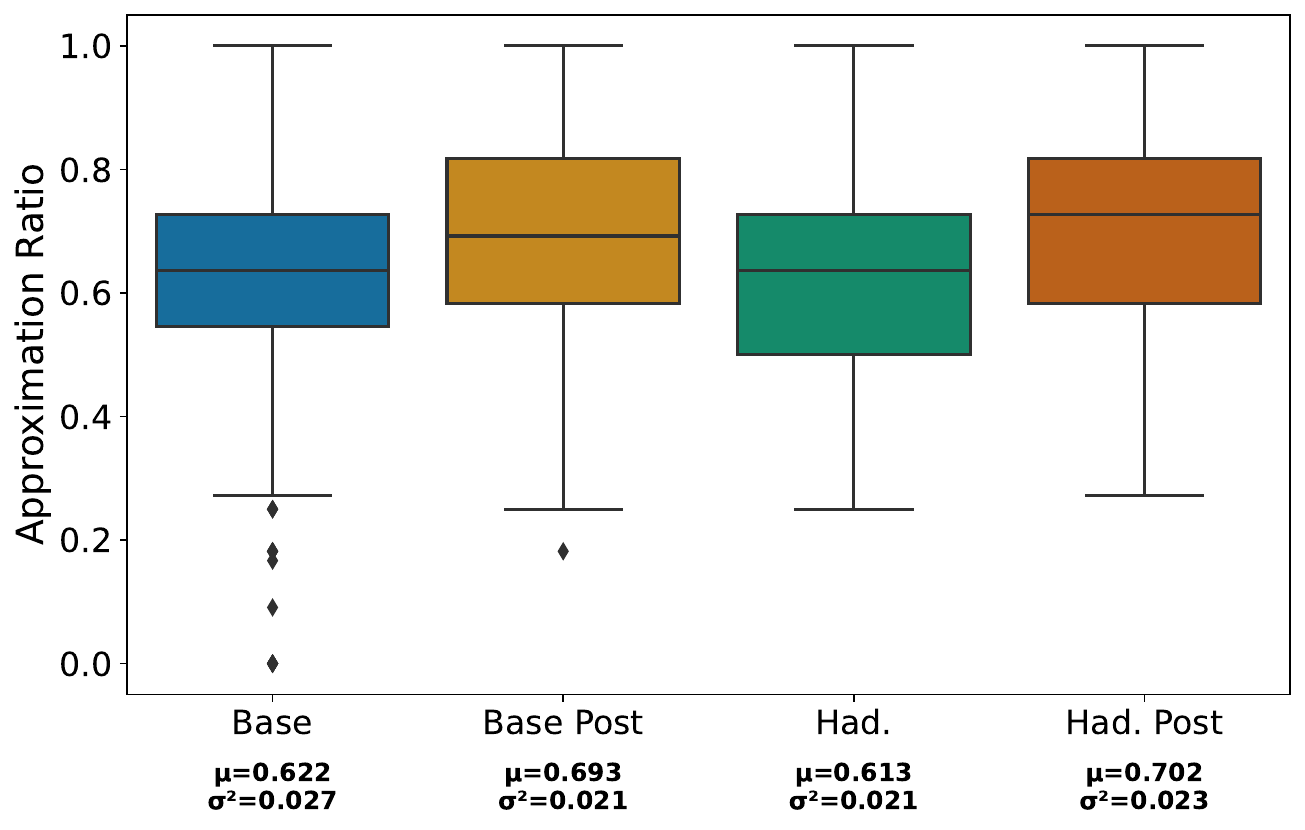}
    \caption{Approximation Ratios of the accelerated VQE Implementations (\texttt{maxiter = 50}). This figure displays the approximation ratios for all four versions of the accelerated VQE setup tested. High values indicate better solution quality, with the best achievable ratio being one.}
    \label{fig:boxplot_approRatio_50}
\end{figure}

\begin{figure*}[!t]
\centering
\includegraphics[width=\textwidth]{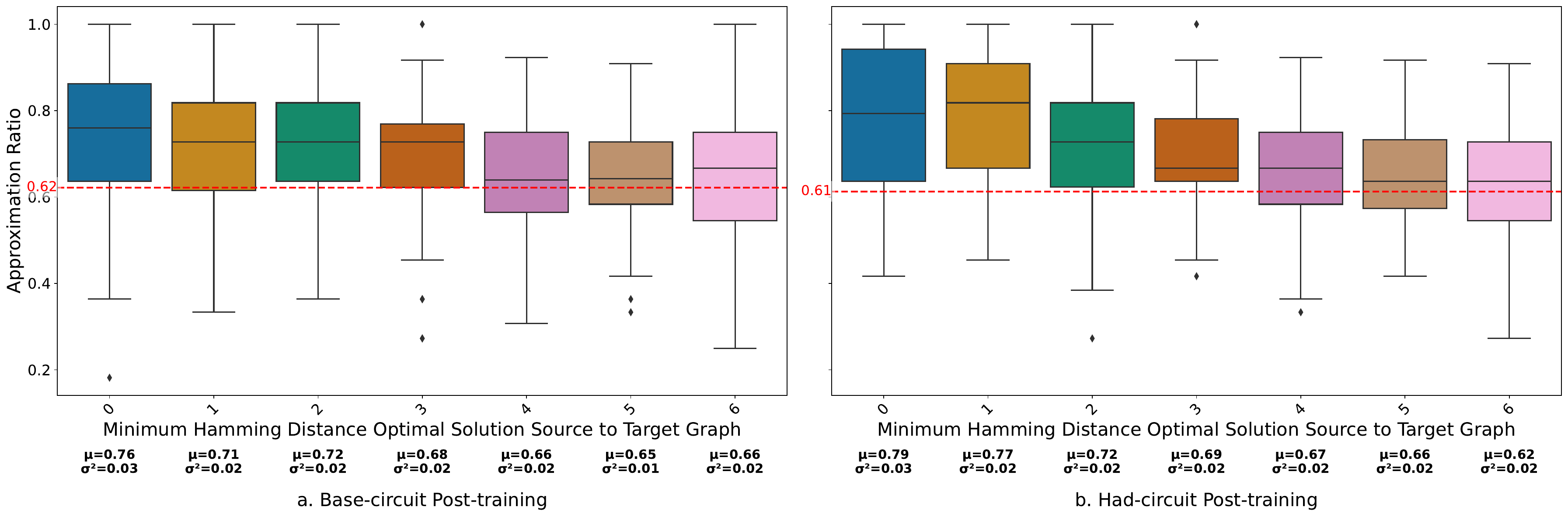}
\caption{Approximation Ratios of the Post-trained accelerated VQE instances with regards to the Solutions HD (\texttt{maxiter = 50}). The figure illustrates the performance of the post-trained accelerated VQE instances (base-circuit left, had-circuit right) with regards to the minimum HD of the optimal solutions of the underlying source-target graph pairs. High values indicate better solution quality, with the best achievable ratio being one.}
\label{fig:boxplot_hd_approRatio_50}
\end{figure*}

We have observed rapid convergence behaviour in the early training phase of the TL approach, as well as improved approximation ratios for problem instances closely related to those used in training. This has prompted the development of an accelerated training for the VQE, which is not a novel algorithm per se, but rather a specialised application example. The frequency at which problem instances occur can be relatively high, potentially preventing practitioners from conducting extensive optimisations due to the need for solutions in a brief timeframe. For the accelerated VQE we have limited the \texttt{maxiter} parameter of the algorithm for all types to 50 iterations (previously 1000), amounting to a reduction of $95\%$ of the training time. With a reduction to $50$, the algorithm shows signs that the parameters are being changed, but complete convergence is not possible. However, the parameters on the source-graph, which are subsequently reused, get still trained by a maximum of $1000$ iterations. \\
Examining the convergence behaviour of the executed algorithms in Figure \ref{fig:combined_convergence_50}, we do not observe the typical behaviour where the algorithm is trained to full convergence. For the base circuit, the TL approach begins at a significantly lower level than the standard initialisation approach and converges in a gradual manner, indicating that further convergence could be expected if not through the parameter setup restricted. The base-circuit with classical parameter initialisation, exhibits a typical convergence behaviour throughout, although the rate of convergence significantly slows in the latter half of the plot. Shifting focus to the right graph, where the had-circuit is depicted, a less stable convergence process is observed. Here again, the expected pattern emerges, with the TL approach starting at a lower level and converging gradually. The classical initialisation approach remains stable until the 30th evaluation, after which it rapidly changes. Both graphs suggest a lower energy level for the TL approach, which we will investigate further in the approximation ratios obtained. \\
To better evaluate the results and relative performance, it is also necessary to consider the corresponding approximation ratios, shown in Figure \ref{fig:boxplot_approRatio_50}. Contrary to previous results, we now observe superior approximation ratios for the TL approach for both circuits. The TL approach achieves an around seven to nine percentage points higher approximation ratio compared to the standard initialisation approach. Across all $700$ instances solved, in an instance by instance comparison, the TL approach outperforms the classical training variant in $55.8\%$ of all instances, achieves comparable results in $15.6\%$, and under-performs in $28.6\%$ of all instances (for the had-circuit $58.1\%$, $16.1\%$ and $25.7\%$ respectively). However, this outcome is tempered by the fact that the approximation levels, compared to the \texttt{maxiter = 1000} configuration, are significantly lower. The base-circuit loses around $20.4$ percentage points, and the had-circuit around $27.7$ percentage points, compared to full convergence. The reason our post-trained circuits now perform relatively better is due to the comparatively smaller loss in approximation quality of approximately $12.9$ and $16.4$ percentage points in the approximation ratio, respectively. \\
Finally, we also analyse in detail the development of solution quality relative to the distance between the optimal solutions of the source- and target-graph (see Figure \ref{fig:boxplot_hd_approRatio_50}). Once again, the data is illustrated in the same manner as before. This time, the average approximation for the classically initialised base-circuits was $0.62$, and for the had-circuit, it was $0.61$. For both circuits, across every HD in optimal solutions, higher average approximation ratios are achieved with our TL approach. Here too, we observe the trend of decreasing solution quality with increasing distance in the optimal solution partitions of the source- and target-graph, further substantiating our results.

\section{Discussion \& Limitations}\label{sec:discussionlimitations}

\subsection{Transfer Learning in General}
Our research demonstrates that initialising VQE parameters with pre-optimised values accelerates early optimisation phases and achieves comparable solution quality upon full convergence. Analysing the distance between optimal solutions of source and target graphs, we found that it´s proximity significantly impacts training success, corroborating findings by Galda et al. ~\cite{galda2023similarity}. Close initial parameters enhance solution quality and convergence, confirming the applicability of TL to VQE and combinatorial optimisation. However, the extent to which retraining affects the parameters and solution quality post-initialisation remains unclear. We identify this as an area for future research to better understand and improve TL techniques. Additionally, expanding this research to weighted MaxCut problems and other combinatorial challenges could further validate the robustness of the TL approach, as suggested by Sureshbabu et al. ~\cite{sureshbabu2024parameter}.

\subsection{Accelerated VQE}
In our accelerated VQE approach, we apply principles from TL research, notably using the rapid early-phase convergence by setting a reduced \texttt{maxiter} parameter, which limits full convergence but starts near the optimal solution. Despite initial doubts about its practicality, the accelerated VQE offers substantial benefits: it reduces circuit evaluations by 93\%, significantly lowering resource use, a figure derived from our experiments. This reduction doesn't achieve the 95\% reduction in \texttt{maxiter} due to early termination upon meeting convergence criteria.\\
While there is a noticeable decline in solution quality due to fewer iterations, we interpret this as a favourable trade-off between computational effort and solution quality, especially compared to standard initialisation methods. This trade-off offers considerable time and cost savings, making it viable for scenarios needing rapid solutions like last-mile delivery. We believe further enhancements in this trade-off can make our method even more appealing.\\
Extensive parameter optimization on the source-graph (\texttt{maxiter = 1000}) is justified when reusing parameters, suggesting potential cost-effectiveness in repeated applications or evolving problem instances. However, our study does not thoroughly examine scalability due to the limited problem sizes tested, and the gap between source and target graphs remains a concern, as it may lead to suboptimal local minima.\\
Additionally, our data hint at the potential for defining universal parameter initialisation for various graph structures. Early results, indicate that TL-initialised circuits start at a lower convergence point, offering an inherent advantage in final approximation ratios. This suggests that pre-training might establish a generally favourable state for similar graphs, proposing a robust starting point for diverse problems. This hypothesis needs validation across broader applications to confirm its efficacy.

\section{Conclusion}\label{sec:conclusion}

We investigated the technique of TL for the VQE algorithm applied to the well-studied MaxCut problem. We have shown that TL also works for this particular combinatorial optimisation problem, as convergence behaviour and solution quality are comparable. Additionally, we have shown that the distance between source and target graphs' optimal solutions, representing the distance to overcome from the starting point of parameter initialisation to the new solution, plays a significant role in the success of the approach execution. A closer optimal solution thereby leads, in general, to better solutions. We have developed the accelerated VQE approach and are able to stop parameter optimisation after only 50 iterations, effectively reducing the resource consumption after necessary pre-training by around $93\%$. Compared to traditional parameter initialisation, the solution quality of the pre-trained parameter initialisation is seven, respectively nine percentage points higher, giving the TL approach a clear advantage. The proposed technique is particularly suited for problem instances which recur in a similar manner. Also, for use cases where a good solution must be reached quickly and/or inexpensively, this approach might be suitable, thus covering many relevant practical applications. Future research should now investigate the solution adaptation between source and target solutions of the TL approach as well as the scaling characteristics of such a technique with respect to the size of the graph and the variability between these graphs.

\section*{\uppercase{Acknowledgements}}
This paper was partially funded by the German Federal Ministry of Education and Research through the funding program “quantum technologies -- from basic research to market” (contract number: 13N16196). J.S. acknowledges support from the German Federal Ministry for Economic Affairs and Climate Action through the funding program “Quantum Computing -- Applications for the industry” based on the allowance “Development of digital technologies” (contract number: 01MQ22008A).

\bibliographystyle{apalike}
{\small
\bibliography{references}}

\newpage

\begin{appendices}
\section*{Appendix}  
\addcontentsline{toc}{section}{Appendix}  

\renewcommand{\thesection}{\Alph{section}}  
\renewcommand{\thesubsection}{\Alph{subsection}}  

\section{Random VQE Initialisation} \label{appendix_a}

\begin{figure}[!h]
\centering
\includegraphics[width=0.49\textwidth]{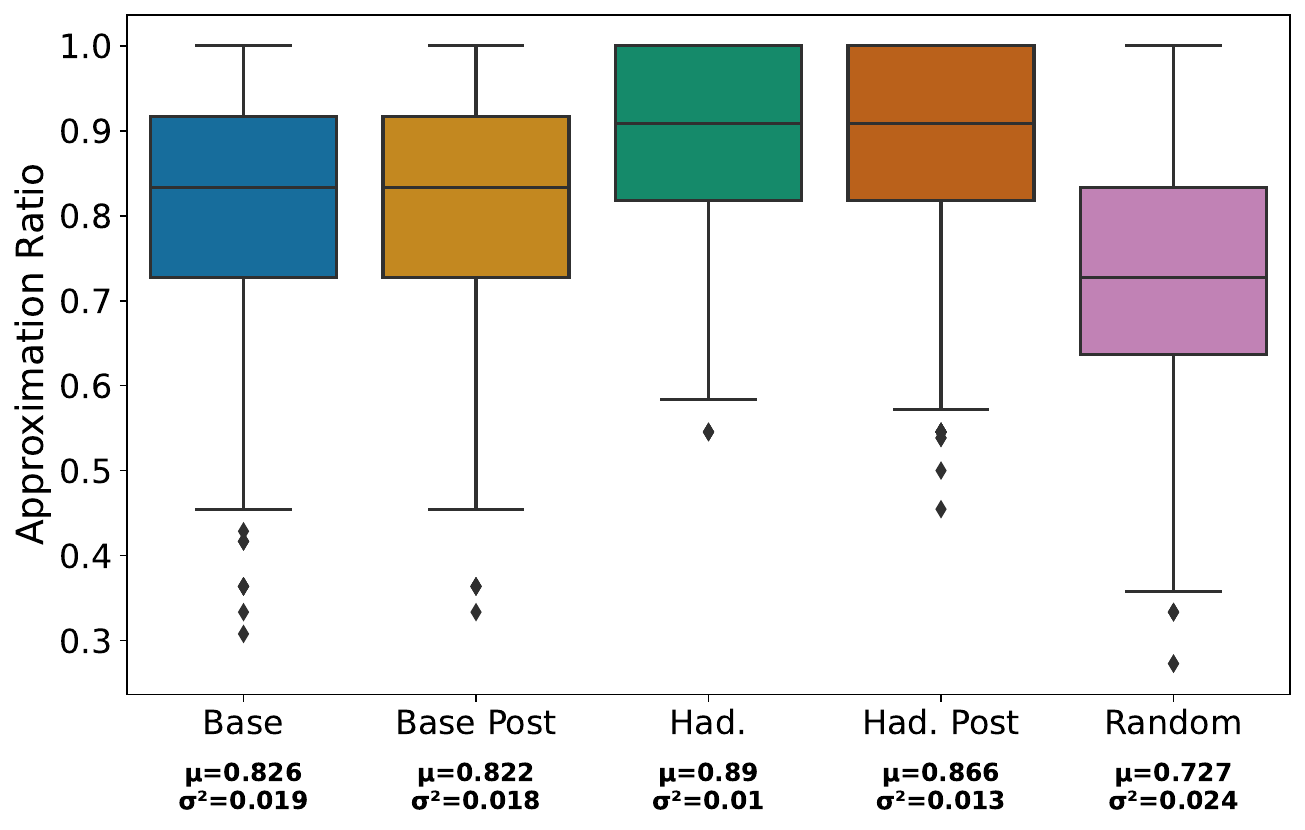}
\caption{Approximation Ratios of VQE Implementations with Random Initialised Instance (\texttt{maxiter = 1000}). This figure displays the approximation ratios for all four main-versions of the VQE tested. Additionally, a random initialised VQE instance was included. High values indicate better solution quality, with the maximum achievable ratio being one.}
\label{fig:boxplot_approxRatios_withRand}
\end{figure}

We have also introduced a version with randomly initialised gate parameters. The objective was to control for the effect of gates being initialised with values not close to zero. The circuit employed is the base-circuit, as hadamard-gates are not required in this configuration, given that gates are already rotated through random initialisation. We observed significantly inferior results, manifested in a lower average approximation ratio and increased variance in these values. This suggests that a random initialisation does not offer any benefits. Further investigation of this setup was not pursued here.

\section{Convergence Behaviour with Regards to HD} \label{appendix_b}

\begin{figure}[!h]
\centering
\includegraphics[width=0.49\textwidth]{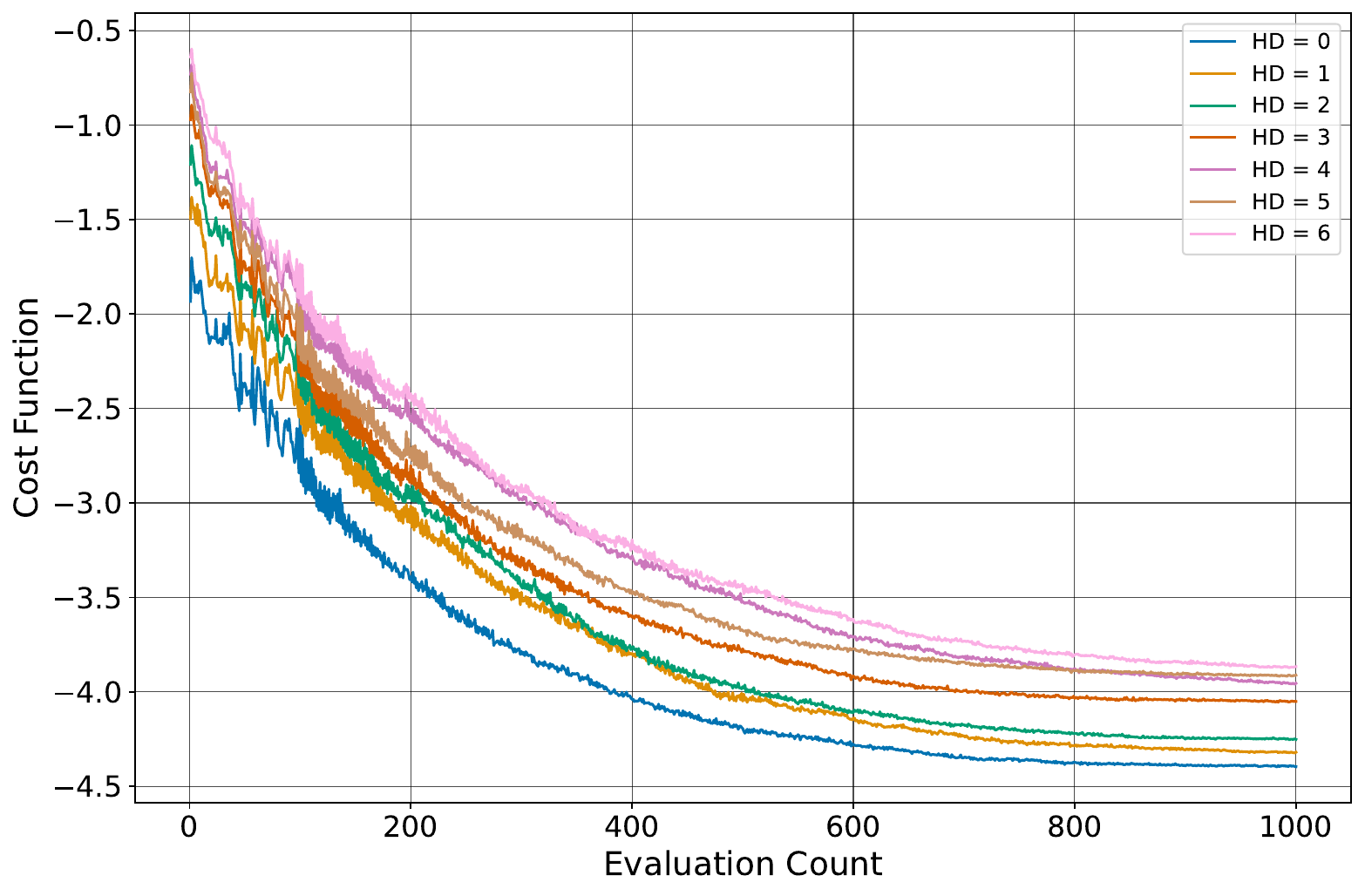}
\caption{Convergence of the Base-VQE Implementation depending on HD (\texttt{maxiter = 1000}). The figure shows the convergence of the base-circuit as a function of the minimum HD between the source- and target-graphs. Lower values are better.}
\label{fig:boxplot_approxRatios_withRand}
\end{figure}

We further assessed the impact of the minimum HD between the source- and target-graphs' optimal partitions on the convergence behaviour. For the sake of clarity, variances have not been presented here. We observe that with increasing HD values, the convergence regarding all three evaluation metrics deteriorates. Specifically, convergence begins at higher values, its slope is flatter, and it reaches a higher level at convergence (indicating poorer performance). The same applies for the had-circuit (not shown here). 
\end{appendices}

\end{document}